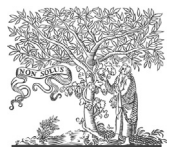

Contents lists available at ScienceDirect

# Journal of Informetrics

journal homepage: www.elsevier.com/locate/joi

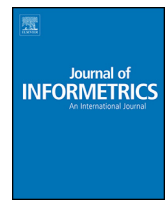

# Is culture related to strong science? An empirical investigation

Mahmood Khosrowjerdi [a],[*], Lutz Bornmann [b]

[a] *Research Department, Inland Norway University of Applied Sciences, Postboks 400, 2418, Elverum, Norway*
[b] *Science Policy and Strategy Department, Administrative Headquarters of the Max Planck Society, Hofgartenstr. 8, 80539, Munich, Germany*

ABSTRACT

National culture is among those societal factors which could influence research and innovation activities. In this study, we investigated the associations of two national culture models with citation impact of nations (measured by the proportion of papers belonging to the 10 % and 1 % most cited papers in the corresponding fields, $PP_{top10\%}$ and $PP_{top1\%}$). Bivariate statistical analyses showed that of six Hofstede's national culture dimensions (HNCD), uncertainty avoidance and power distance had a statistically significant negative association, while individualism and indulgence had a statistically significant positive association with both citation impact indicators ($PP_{top10\%}$ and $PP_{top1\%}$). The study also revealed that of two Inglehart-Welzel cultural values (IWCV), the value survival versus self-expression is statistically significantly related to both citation impact indicators ($PP_{top10\%}$ and $PP_{top1\%}$). We additionally calculated multiple regression analyses controlling for the possible effects of confounding factors including national self-citations, international co-authorships, investments in research and development, international migrant stock, number of researchers of each nation, language, and productivity. The results revealed that the statistically significant associations of HNCD with citation impact indicators disappeared. But the statistically significant relationship between survivals versus self-expression values and both citation impact indicators remained stable even after controlling for the confounding variables. Thus, the freedom of expression and trust in society might contribute to better scholarly communication systems, higher level of international collaborations, and further quality research.

## 1. Introduction

. . . scientists do not just collect data, they design experiments to collect the data and they refine and interpret both the data and experiments. In each case what they could do and what they actually do are influenced by their motivations, by the social, informational, and technological resources available, and by the available alternative theories and models. Scientists also write, read, and argue about data, models, theories, and explanations, and in each case there are social and cultural contexts that influence their interpretations or their choice of statements. (Duschl et al., 2006)

* Corresponding author.
  *E-mail address:* mahmood.khosrowjerdi@inn.no (M. Khosrowjerdi).

Prior research has shown that various economic, demographic, and policy-related factors could influence the scientific performance of nations (measured in terms of citations and other indicators). Of the economic factors, the wealth of nations (Baker et al., 2015; Ertekin, 2014; Gantman, 2012; Monge-Nájera & Nielsen, 2005; Mueller, 2016; Shukla & Bauer, 2012) and the allocated expenditures to research and development (Baker et al., 2015; Erfanian & Ferreira Neto, 2017; Mueller, 2016; Shukla & Bauer, 2012) have been reported as predictors of national performance. The population (Monge-Nájera & Nielsen, 2005), the size of national publication industry (Mueller, 2016), the number of university graduates/researchers in a country (Erfanian & Ferreira Neto, 2017; Shukla & Bauer, 2012), and (English) language proficiency (Gantman, 2012) are among the demographic predictors of scientific performance. The policies and regulations of nations (Harzing & Giroud, 2014), the science culture (Ertekin, 2014; Inönü, 2003; Shukla & Bauer, 2012), the free mobility of researchers (Sugimoto et al., 2017), and the general academic openness of nations (Wagner & Jonkers, 2017) seem to be related to national scientific performance as well.

There is research evidence that systems of thought of individuals are culturally shaped, for example, researchers (Nisbett et al., 2001) revealed that the East Asians had relatively holistic mode of thinking (i.e. identifying large-scale patterns and responding to them) while Westerns had analytic mode of thinking (i.e. focusing on an object and its associated categories, and using logic to explain the behavior of the object). Furthermore, thinking styles/modes could influence teaching and learning outcomes (Cheng et al., 2011) and especially, academic performance of individuals (Grigorenko & Sternberg, 1997; Zhang, 2002). However, the relationship between culture and national scientific performance is not well researched, and the limited number of publications in this domain is not conclusive. This understanding is important since it is the first step to a fuller explication of the cultural differences/similarities that could contribute to scientific performance of nations.

National culture is the shared, collective values that differentiates one group of individuals from another ((Hofstede, 2011). National culture is among those societal factors which could influence research and innovation activities (Couto & Vieira, 2004; Jones & Davis, 2000; Yair, 2020) and productivity (Kedia et al., 1992). In our search of the literature, we found only three studies which focused on the correlation between national culture and citation impact, although citation impact is one of the most important metrics to measure the research performance of individuals, groups of researchers, institutions or world nations. For example, the study by Abraham (2020) showed that individualist and low-hierarchy nations are more likely to have higher citation impact comparing with collectivist nations in which the hierarchical societal systems are prevalent. The three studies provided some insights into the relationship of national culture and citation impact, but are concerned by several limitations. For example, the citation impact of nations may be resultant to many factors such as allocated expenditures to research and development, number of researchers of a country, and the level of national self-citations (Waltman, 2016). These factors should be controlled in the statistical analyses of the relationship between impact and culture. Other limitations refer to the absence of robustness tests and the use of inappropriate indicators for measuring citation impact.

In this study, therefore, we applied two separate cultural models and the relevant datasets, that is, Hofstede's national culture dimensions (HNCD) and Inglehart-Welzel cultural values (IWCV). We linked the models with advanced citation impact indicators. We estimated the associations of national culture dimensions with citation impact of nations after controlling for country-level data on openness of nations (international migrant stock), academic openness (international co-authorships), national self-citations, number of researchers, investments in research and development, language, and productivity.

## 2. Literature review

### 2.1. National culture frameworks

This research used two cultural frameworks to explain the relationships of national culture and the citation impact of nations. The first (Hofstede et al., 2010) perceives national culture as a stable construct which does not change in short term periods (or it changes in very long time periods), and the second (Inglehart, 1997) assumes the dynamics of national culture during time. Recent research (Beugelsdijk & Welzel, 2018; Beugelsdijk et al., 2015) showed that both approaches are complementary, and although young generations seem to become more individualistic (which supports the dynamic approach), one can assume that nearly half of national culture values are stable.

This research investigates the possible connections of noted cultural frameworks with citation impact of nations. This helps us to identify those cultural dimensions which could influence citation impact (if any). However, the two applied cultural frameworks and their associated dimensions are based on different theoretic assumptions and could not be used interchangeably.

#### 2.1.1. Hofstede's national culture dimensions (HNCD)

Hofstede et al. (2010) categorized the world countries into six bi-polar dimensions of national culture: small versus large power distance, individualism versus collectivism, masculinity versus femininity, low versus high uncertainty avoidance, short- versus long-term orientations, and indulgence versus restraint. As Hofstede (2011) explains:

- The power distance shows the degree of acceptance of power inequality among the individuals in lower classes (pp. 61–62) and it is manifested in interactions of people in family (parent-children), school (teacher-students), working place (manager-employee), and country (authorities-citizens). In power distant nations (those which have large power distance), the education system is hierarchical and injective which means the educators are regarded as unquestioned authorities (pp. 67–79).
- The individualism-collectivism dimension is the extent of group orientations among people in a country. In individualist nations, the people are supposed to take care of themselves and their close family members, while in collectivist nations, the people are bounded by in-group loyalties (pp. 92–93).
- The masculinity-femininity shows the degree in which nations have overlapped or separated gender roles and values (p. 140).
- The uncertainty avoidance refers to the capacity of individuals in a country to recover quickly from vague or ambiguous situations (p. 191).
- The perceptions of people towards the time and their following behaviors shape the time orientation dimension (pp. 237–239).
- The indulgence versus restraint dimension refers to the degree of freeness or regulation of gratification in a country (p. 281).

*2.1.2. Inglehart-Welzel cultural values (IWCV)*

Inglehart and Welzel (1984–2015) analyzed the world values survey (WVS) data and categorized world nations into two bipolar dimensions of culture, that is, 1) traditional versus secular-rational values and (2) survival versus self-expression values. Inglehart and Welzel (n.d.) clustered the world nations based on the proximity of cultural values, and grouped, for example, United States, Canada, Australia, and United Kingdom in the same group. The researchers defined the first dimension of culture based on the degree of importance of religion, traditions, and obedience of authorities. The world nations were categorized into a bipolar spectrum of traditional nations on the one hand in which the religion, family ties, obedience of authorities are very important, such as Islamic-African countries. On the other hand are the secular-rational nations such as protestant European nations in which the religion, family ties, obedience of authorities are less important. Inglehart and Welzel (n.d.) characterized the second bipolar dimension of culture according to the emphasis of nations on economic and personal security at the one end of the spectrum (i.e. nations with survival values, such as most of African and South Asian countries) and the freedom of expression and active participation in social and political decision-making at the other end (i.e. nations with self-expression values, such as Scandinavian countries).

According to Inglehart and Welzel (2005, p. 5), national culture is a dynamic phenomenon and subject to change, and the socioeconomic development plays a major role in the change process. Inglehart and Welzel (2005, p. 5) claimed that industrial and postindustrial modernization episodes change the cultural values in a certain way: a move from conventional to secular-rational beliefs during the industrial revolution, and a move from survival to self-expression beliefs during the postindustrial period which contributed to increased independence from authorities. This thinking in episodes is different from Hofstede's approach which assumes national culture as a stable construct.

*2.2. Factors influencing citation impact*

*2.2.1. Cultural factors*

Previous studies (Almeida et al., 2009) show that geographic proximity and cultural commonality influence science communication patterns of world nations. Jiang et al. (2018) revealed that co-authored publications from nations with similar economies and different political settings had received higher citations in contrast to those from nations with different economies and similar political settings. A selective analysis of citations to first-author females in five countries (Turkey, India, Spain, UK, and the USA) (Thelwall, 2018) demonstrated that the first-author females were less cited in Turkey and India, but slightly more cited in Spain, UK, and the USA. The author mentioned "gender bias" in those two nations as a possible explanation of this outcome. Abbasi and Jaafari (2013) showed that co-authorships on all levels (institutional, national, and international) are statically significantly correlated with the number of received citations of the corresponding papers. The study also worked out that international co-authorships are correlated on a lower level with the number of citations than other types of co-authorships (institutional and national). The authors highlighted cultural/cross-national challenges in international collaborations as a likely justification for this finding.

In our search of the literature, we found only a few studies that have focused on the direct correlation between cultural values and national academic performance. These studies have been published in recent years.

Andrijauskienė and Dumčiuvienė (2017) studied the associations of national culture (based on HNCD) and the nations' innovativeness (based on data extracted from the Summary Innovation Index – a sub-index of the European Innovation Scoreboard) for 27 EU member states. The authors showed that of the HNCD, the dimensions of indulgence and individualism were positively while power distance and uncertainty avoidance were negatively correlated with national innovation performance. Further regression analyses revealed that just two dimensions of power distance and indulgence were statistically significant predictors of nations' innovativeness.

The study by Allik et al. (2020) showed that after controlling for variables of wealth, i.e. research and development expenditures, economic inequality, population, and the worldwide governance indicator (WGI), the "good governance" (measured by WGI) was a strong predictor of the scientific impact of nations. The WGI included six factors of responsibility and effectivity of government, degree of violence, the ability of the government to formulate and implement sound policies and regulations, the authority and influence of law in society, and the corruption level of nations. Allik et al. (2020) measured the scientific impact of nations by three indicators: 1) the nations' average citation rate per paper, 2) the proportion of papers of a nation within the top 1% most cited publications, and 3) the number of research categories in which a country is listed in the Essential Science Indicators (ESI).

Abraham (2020) analyzed HNCD data and SCImago Journal & Country Rank data for 60 countries and showed that of six dimensions of HNCD, three dimensions of power distance, individualism, and indulgence had statistically significant correlations with the research impact of nations (measured by the number of published documents, citable documents, citations, self-citations, h index values, and citations per document for countries). After controlling for the wealth of countries, Abraham (2020) found that uncertainty avoidance was (negatively) correlated with the research impact of nations, too.

### 2.2.2. Confounding factors

Earlier studies indicated that institutional and national self-citations could bias the assessment of institutions and nations by using citation impact indicators. For instance, Aksnes (2003) analyzed 45,000 publications (articles, notes, reviews, and proceeding papers) of Norway (for the years 1981–1996) and revealed that self-citations on the author level constituted 36 % of all received citations. This finding was confirmed in a later study with a larger sample (Fowler & Aksnes, 2007). Glänzel et al. (2004) revealed that self-citations were prevalent for 50 studied countries (varied from 53.14 % for the Ukraine to 22.10 % for the United States). Highly developed nations such as the United States, Canada, and the United Kingdom had the lowest self-citation rate.

National investments in research and development have been reported as a predictor of the quantity of publications of nations (Baker et al., 2015; Wagner & Jonkers, 2017; Wagner et al., 2018). However, the evidence on the effects of those investments on the quality (citation impact) of publications is limited. Recent studies (Wagner & Jonkers, 2017; Wagner et al., 2018) have shown that the national investments in research and development are not a statistically significant predictor of citation impact of nations, but the openness of nations (measured by international co-authorships and free mobility of scholars). In other words, the international "engagements" of nations attract the attention to their research system (Wagner & Jonkers, 2017).

Some studies have shown that the quantity (i.e. the number of publications) (Lee & Bozeman, 2005) and the quality (received citations) of academic contributions of researchers (Gazni & Didegah, 2011) are positively associated with co-authorships. For example, Li et al. (2013) analyzed the academic productions of Information Systems scholars (extracted from the Social Science Citation Index database, Clarivate Analytics). The authors showed that those scholars who were connectors in co-authorship networks were more likely to produce publications with higher citation counts. This co-authorship effect is also visible on the aggregated national level.

In a recent study, Wagner et al. (2018) studied the role of national openness (international co-authorships and mobility of scholars) in the context of national performance. The researchers operationalized performance as fractionally field-weighted citation impact (FWCI) of nations. Colledge (2014) defines the FWCI as "the ratio of citations received (by a nation) relative to the expected world average for the subject field, publication type, and publication year". Controlling for national investments in research and development in the statistical analyses, Wagner et al. (2018) showed that academic openness of nations was a strong predictor of national citation impact.

There is some evidence in the literature that the number of researchers in a country is a predicting factor of national research outcome. For instance, Erfanian and Ferreira Neto (2017) used two nation-level datasets (from the World Bank and Organization for Economic Cooperation and Development, OECD) to explore the determining factors of 31 nations' research outcomes. The authors found for the OECD dataset that the number of researchers was a stronger predictor of national research outcome than capital. However, the analysis of the World Bank dataset revealed that capital was the most important predictor of research outcome. The effect of the number of researchers on national research outcome could be shown already in an earlier study by Shukla and Bauer (2012). In this study, the number of new sciene and engineering (S&T) graduates was a positive predictor of the national science and technology performance.

Some researchers investigated possible links between national/primary language of authors, language of papers, and received citations of papers. For instance, Borsuk et al. (2009) did not find any correlation between primary language of first authors of papers in ecology and citation rates. Diekhoff et al. (2013) found that the higher number of English-language articles published in multi-lingual journals and the journals' country of origian were correlated with higher impact factor (IF) scores; and the North American journals had higher IF than journals origirating from Europe or South America. Bornmann et al. (2012) revealed that the (English) language of pulishing journals was correlated with the number of received citations.

Researchers have revealed the association of productivity (i.e. published papers of a project) and citation rates (e.g. Bornmann & Daniel, 2007), and there is some evidence that the size of national publication industry influence the scientific performance of nations measured in terms of number of citable documents (Mueller, 2016).

### 2.3. Rationale for this study

While the current literature provides some evidence for the relationships of national culture dimensions/values and citation impact, the results are not inclusive. The studies are concerned by several limitations which we tried to avoid in this study. First, we used two different cultural approaches (HNCD and IWCV) to investigate the relationship of national culture and citation impact. We wanted to test the robustness of the results. Second, we went beyond bivariate analyses and used advanced regression models to analyze the data. Third, we used two advanced citation impact indicators, i.e. the proportions of papers belonging to the 10 % and 1 % most frequently cited papers in the corresponding subject category and publication year ($PP_{top10\%}$ and $PP_{top1\%}$). Fourth, we undertook parallel statistical analyses (i.e. correlation analysis, linear regression analysis, and fractional logistics regression analysis) to investigate the stability of associations between national culture dimensions/values and citation impact. Fifth, we controlled for the effects of many possible confounding variables: national self-citations, international co-authorships, investments in research and development, international migrant stock, number of researchers of each nation, language, and productivity.

## 3. Methods

### 3.1. Cultural datasets

This research used two nation-level cultural datasets to investigate the relationships of national culture and the citation impact of nations.

#### 3.1.1. HNCD data
The HNCD data comprises nation-level scores of six cultural dimensions of small versus large power distance, individualism versus collectivism, masculinity versus femininity, low versus high uncertainty avoidance, short- versus long-term orientations, and indulgence versus restraint. This data was available for 112 nations.

#### 3.1.2. IWCV data
The IWCV data includes national-level scores for two cultural dimensions of 1) traditional versus secular-rational values and (2) survival versus self-expression values. The data was accessible for 66 nations.

### 3.2. Citation impact data

The citation impact data were exported from the Max Planck Society's in-house database which is based on the Web of Science (Clarivate Analytics). The data included the national proportions of papers (articles and reviews published in 2006 and 2010) belonging to the top 1% ($PP_{top\ 1\%}$) and top 10 % ($PP_{top10\ \%}$). $PP_{top10\ \%}$ is the proportion of papers (published by a country) which belong to the 10 % most frequently cited papers within their subject categories and publication years (correspondingly, $PP_{top\ 1\%}$ refers to the 1% most frequently cited papers) (Hicks et al., 2015). We used fractional instead of full counting of papers: the national numbers of (highly cited) papers were weighted by the number of countries on a paper. For example, if there were three countries mentioned on a paper, each country received 1/3 of the paper.

### 3.3. Confounding variables

In order to investigate the association of cultural dimensions/values and citation impact of nations in this study, we controlled for the possible effects of several confounding variables. These variables have been revealed as (possible) influencing factors of citations (on the national level) in the scientometric literature. Based on the results of former studies, we considered several variables in the current study that might confound the relationship of cultural values and citation impact such as national self-citations (McGarty, 2000), international co-authorships (Glänzel, 2001; Li et al., 2013; Wagner et al., 2018), investments in research and development (Baker et al., 2015; Wagner & Jonkers, 2017; Wagner et al., 2018), international migrant stock (as an indicator of nations' openness) (Wagner & Jonkers, 2017), and the number of researchers (Shukla & Bauer, 2012; Erfanian & Ferreira Neto, 2017), language (Bornmann et al., 2012; Diekhoff et al., 2013), and productivity (Bornmann & Daniel, 2007). The detailed description of the datasets for confounding variables and the sources are presented in Table 1.

### 3.4. Overview of the datasets

The IWCV dataset was available for the year 2006, and the last update of HNCD dataset was for the year 2010. To align the citation impact data with the cultural datasets, years 2006 and 2010 were selected as reference points in this study. For citation impact indicators, only those nations which had published ≥500 articles and reviews are included in the analyses. The alignment of the two cultural datasets (IWCV and HNCD) with the citation impact indicators resulted in different samples in the reported statistical analyses of this study, that is, 36 for the IWCD dataset, and 58 for the HNCD dataset.

Aligning the existing data of confounding variables with the cultural and citation impact data reduced the final samples to 28 (IWCV dataset) and 44 (HNCD dataset) nations. The applied datasets, their descriptions, and original sources are summarized in Table 1.

### 3.5. Statistics

This study used several statistical methods to explore the relationships of cultural dimensions (as independent variables) with citation impact indicators (as dependent variables) by controlling for seven confounding variables: national self-citations, international co-authorships, investments in research and development, international migrant stock, number of researchers of each nation, language, and productivity. Since the dependent variables of this study (i.e. citation impact) are proportions (see above), we computed fractional logistic regression models using STATA 16.0. The fractional logistic regression (using the *fracreg* command) suits for a dependent variable (such as rates, proportions, and fractional data) that has values in the range of $0 \leq$ values $\leq 1$. Various models (probit, logit, or heteroskedastic probit) can be computed for the conditional mean (STATA, n.d.) which usually lead to similar results. In this study, we computed the logit model because of its relatively simple interpretation (using the *fracreg logit* command).

**Table 1**
Description of the datasets.

| Variable(s) | | Description | Source | Year |
| --- | --- | --- | --- | --- |
| Culture | HNCD dataset | The nation-level scores of six cultural dimensions of small versus large power distance, individualism versus collectivism, masculinity versus femininity, low versus high uncertainty avoidance, short- versus long-term orientations, and indulgence versus restraint | Hofstede's website[a] | 2010 |
| | IWCV dataset | The nation-level scores of two cultural dimensions of traditional versus secular-rational values, and survival versus self-expression values | World Values Survey website[b] | 2006 |
| Citation impact | $PP_{top\ 10\%}$ | The national proportions of papers (articles and reviews published in 2006 and 2010) which belong to the 10 % most frequently cited papers within their subject categories and publication years | The data were exported from the Max Planck Society's in-house database which is based on the Web of Science (Clarivate Analytics) | 2006 and 2010 |
| | $PP_{top\ 1\%}$ | The national proportions of papers (articles and reviews published in 2006 and 2010) which belong to the 1% most frequently cited papers within their subject categories and publication years | The data were exported from the Max Planck Society's in-house database which is based on the Web of Science (Clarivate Analytics) | 2006 and 2010 |
| Control variables | National self-citation | Proportions of papers by a nation citing another publication of the same nation | The data were exported from the Max Planck Society's in-house database which is based on the Web of Science (Clarivate Analytics) | 2006 and 2010 |
| | International co-authorships | Proportions of co-authored papers by nations with international researchers | The data were exported from the Max Planck Society's in-house database which is based on the Web of Science (Clarivate Analytics) | 2006 and 2010 |
| | Investments in research and development | Proportions of national investments in research and development (% GDP) | World Bank | 2006 and 2010 |
| | International migrant stock[c] | International migrant stock (% of population) | World Bank, World development indicators, indicator code: SM.POP.TOTL.ZS | 2005 and 2010 |
| | Number of researchers | Number of researchers (per million inhabitants) | UNESCO Institute for Statistics, access via World Bank, (indicator code: SP.POP.SCIE.RD.P6) | 2006 and 2010 |
| | Language | Binary variable: whether or not English is the official language in the country | – | – |
| | Productivity | National number of papers | The data were exported from the Max Planck Society's in-house database which is based on the Web of Science (Clarivate Analytics) | 2006 and 2010 |

[a] https://geerthofstede.com/research-and-vsm/dimension-data-matrix/.
[b] https://web.archive.org/web/20131019112321/http://www.worldvaluessurvey.org/wvs/articles/folder_published/article_base_54/files/ValueScores_5_waves.doc.
[c] International migrant stock data was not available for the year 2006, and we used the closest data which was available for the year 2005.

In order to investigate the stability of our results based on fractional logistic regression models, we used IBM SPSS (version 25) to perform linear regression analyses and Pearson's correlation analyses.

James et al. (2013) list the important assumptions of linear regression analyses as linear relationships of predictors and response (by investigating the plots of residuals versus predicted values), outlier-free data ($-3 < std. residual < +3$), and the absence of collinearity. The standard residuals in our study were between the cutoff threshold ($i.e.\ -2.594 < std.\ residual < +2.364$), the variance inflation factor (VIF) of predictor variables had values less than 5, and the tolerance statistics were higher than 0.10 for all variables in this study. These values are in accordance with the cutoff threshold (Craney & Surles, 2002; James et al., 2013).

The results of the stability analyses are part of the supporting information that can be found at https://dataverse.no/dataverse/inn.

## 4. Results

The findings of the fractional regression analyses are presented in this section. The findings show the associations of national culture (measured by two separate datasets of HNCD and IWCV) and two citation impact indicators ($PP_{top\ 1\%}$ and $PP_{top10\%}$) as dependent variables, before and after controlling for confounding variables of national self-citations, international co-authorships, investments in research and development, international migrant stock, number of researchers of each nation, language, and productivity.

### 4.1. HNCD as predictors of citation impact

Table 2 shows the results of the fractional logistic regression analysis for investigating the direct effects of HNCD on citation impact measured by $PP_{top10\%}$. As it is presented in Table 2 (Model 1), two dimensions of power distance and uncertainty avoidance have statistically significant effects on $PP_{top10\%}$. The effects of those dimensions of HNCD on $PP_{top10\%}$ are as follows. A 1 percentage point increase in the power distance score of nations is associated with a 2.9 percentage points decline in $PP_{top10\%}$; and a 1 percentage point increase in the uncertainty avoidance score of nations is associated with a 3.2 percentage points decline in $PP_{top10\%}$. After controlling for the possible effects of the confounding variables (national self-citations, international co-authorships, investments in research and development, international migrant stock, number of researchers of each nation, language, and productivity) none of the HNCD dimensions had a statistically significant effect on $PP_{top10\%}$ (see Table 2, Model 2).

Model 3 in Table 2 depicts the results of the fractional logistic regression analysis for exploring the direct effects of HNCD on citation impact measured by $PP_{top\ 1\%}$. The results show that two dimensions of uncertainty avoidance and individualism had statistically significant effects on $PP_{top\ 1\%}$. The average marginal effects (AME) reveal that a 1 percentage point increase in uncertainty avoidance is associated with a .008 percentage points decline; and a 1 percentage point increase in individualism is associated with a .006 percentage points increase in $PP_{top1\%}$. After controlling for the possible effects of confounding variables (national self-citations, international co-authorships, investments in research and development, international migrant stock, number of researchers of each nation, language, and productivity) none of the HNCD dimensions had a significant effect on $PP_{top\ 1\%}$ (Table 2, Model 4). Thus, the HNCD dimensions could not predict the citation impact of nations (after controlling for the effects of confounding variables).

### 4.2. IWCV as predictors of citation impact

Table 3 shows the results of the fractional logistic regression analyses for investigating the effects of IWCV on citation impact. The results of Model 5 reveal that of two dimensions, only survival versus self-expression values indicated a statistically significant and positive effect on $PP_{top10\%}$ as follows: a 1 percentage point increase in self-expression values of nations was associated with a 2.3 percentage points increase in their citation impact (measured by $PP_{top10\%}$).

After controlling for the possible effects of confounding variables (national self-citations, international co-authorships, investments in research and development, international migrant stock, number of researchers of each nation, language, and productivity) (Table 3, Model 6), the statistically significant effect of self-expression values on $PP_{top10\%}$ remained stable as follows: a 1 percentage point increase in self-expression values was associated with a 1.4 percentage points increase in $PP_{top10\%}$.

Table 3 (Model 7 and Model 8) shows the results of the fractional logistic regression models and average marginal effects for IWCV and $PP_{top\ 1\%}$. The results of Model 7 indicated that just self-survival versus expression values had a statistically significant effect on $PP_{top\ 1\%}$. A 1 percentage point increase in self-expression values of nations was associated with a .28 percentage points increase in $PP_{top\ 1\%}$. After controlling for the possible effects of confounding variables (national self-citations, international co-authorships, investments in research and development, international migrant stock, number of researchers of each nation, language, and productivity) (Table 3, Model 8), the statistically significant effect of self-expression values on $PP_{top\ 1\%}$ remained constant as follows: a 1 percentage point increase in self-expression values was associated with a .18 percentage points increase in $PP_{top\ 1\%}$. Summarizing the results of Table 3, it seems that of the IWCV dimensions, only the value survival versus self-expression was a stable predictor of citation impact of nations. It means that those nations which have predominant survival values in society have low citation impact, while those nations which respect self-expression have high citation impact.

**Table 2**
The HNCD as predictors of citation impact.

| | | Citation impact indicator | | | | | | | |
|---|---|---|---|---|---|---|---|---|---|
| | | $PP_{top\ 10\%}$ (2010) | | | | $PP_{top\ 1\%}$ (2010) | | | |
| | | Model 1 | | Model 2 | | Model 3 | | Model 4 | |
| | | Coefficient | AME | Coefficient | AME | Coefficient | AME | Coefficient | AME |
| Confounding variable | National self-citations (2010) | | | 1.362 (−1.05) | .019768 | | | 1.206 (−0.83) | .0017318 |
| | International co-authorships (2010) | | | 0.0129* (−2.04) | .0420956 | | | 0.0183* (−2.39) | .0061395 |
| | Investments in research and development (2010) | | | 0.102 (−0.84) | .0148766 | | | 0.0876 (−0.65) | .0011397 |
| | International migrant stock (2010) | | | 0.0213*** (−4.11) | .0158584 | | | 0.0287*** (−6.04) | .0024758 |
| | Number of researchers (2010) | | | 0.000023 (−0.39) | .0054059 | | | 0.0000226 (−0.36) | .0005766 |
| | Language | | | −0.227 (−1.31) | .0017205 | | | −0.177 (−1.09) | −.0002255 |
| | Productivity (2010) | | | 0.00000103 (−0.97) | .0032422 | | | 0.00000169 (−1.69) | .0005735 |
| HNCD | Power distance | −0.00784* (−2.02) | −.0297685 | −0.00497 (−1.26) | −.0185004 | −0.00602 (−1.38) | −.0000428 | −0.00543 (−1.28) | −.0019511 |
| | Individualism | 0.00575 (1.71) | .0214057 | 0.0038 (−0.98) | .0151977 | 0.00786* (1.97) | .0000559 | 0.00396 (−0.96) | .0016615 |
| | Masculinity | −0.00128 (−0.62) | −.0042538 | 0.000109 (−0.05) | .0003556 | −0.000265 (−0.11) | −1.88e-06 | 0.000803 (−0.31) | .0002635 |
| | Uncertainty avoidance | −0.00766* (−2.53) | −.0319571 | −0.00288 (−1.08) | −.0121564 | −0.0115** (−2.88) | −.0000819 | −0.00377 (−1.31) | −.0015084 |
| | Long-term orientation | 0.00430 (1.78) | .0143485 | −0.000253 (-0.09) | −.0008935 | 0.00574 (1.82) | .0000408 | 0.00247 (−0.87) | .000885 |
| | Indulgence | 0.00463 (1.52) | .0163688 | 0.00311 (−0.95) | .0108949 | 0.00688 (1.87) | .000049 | 0.00448 (−1.51) | .0016385 |
| (Constant) | | −2.218*** (−4.27) | | −3.686*** (−4.94) | | 4.895*** (−7.98) | | −6.657*** (−8.68) | |
| N | | 58 | | 44 | | 58 | | 44 | |

t statistics in parentheses.
* $p < 0.05$, ** $p < 0.01$, *** $p < 0.001$.
AME: Average marginal effects.
N: number of nations included in the analysis.

**Table 3**

The IWCV as predictors of citation impact.

| | | Citation impact indicator | | | | | | | |
|---|---|---|---|---|---|---|---|---|---|
| | | $PP_{top\ 10\%}$ (2006) | | | | $PP_{top\ 1\%}$ (2006) | | | |
| | | Model 5 | | Model 6 | | Model 7 | | Model 8 | |
| | | Coefficient | AME | Coefficient | AME | Coefficient | AME | Coefficient | AME |
| **Confounding variable** | National self-citations (2006) | | | −0.763 (−0.45) | −.0119358 | | | −0.88 (−0.40) | −.0013379 |
| | International co-authorships (2006) | | | 0.00198 (−0.33) | .0059973 | | | 0.0033 (−0.48) | .0009597 |
| | Investments in research and development (2006) | | | 0.138 (−1.08) | .0158078 | | | −0.00925 (−0.07) | −.0001137 |
| | International migrant stock (2005) | | | −0.00428 (−0.30) | −.0025401 | | | −0.00122 (−0.09) | −.0000754 |
| | Number of researchers (2006) | | | −0.0000792 (−1.05) | −.0176233 | | | −0.000016 (−0.20) | −.0003644 |
| | Language | | | 0.248 (−1.17) | .0038806 | | | 0.264 (−1.2) | .000486 |
| | Productivity (2006) | | | 0.00000191 (−1.2) | .0066272 | | | 0.00000325 (−1.58) | .0012725 |
| **IWCV** | Traditional versus secular-rational values | 0.0661 (1.26) | .0047624 | 0.129 (−1.55) | .002252 | 0.0699 (0.92) | .0004879 | 0.172 (−1.74) | .0003321 |
| | Survival versus self-expression values | 0.319*** (6.87) | .0229785 | 0.270*** (−3.63) | .0142748 | 0.408*** (7.00) | .0028452 | 0.290*** (−3.67) | .0017779 |
| Constant | | −2.633*** (−51.37) | | −2.642*** (−4.83) | | −5.212*** (−89.91) | | −5.267*** (−7.64) | |
| N | | 36 | | 28 | | 36 | | 28 | |

t statistics in parentheses. * p < 0.05, ** p < 0.01, *** p < 0.001.

N: number of nations included in the analysis.

AME: Average marginal effects.

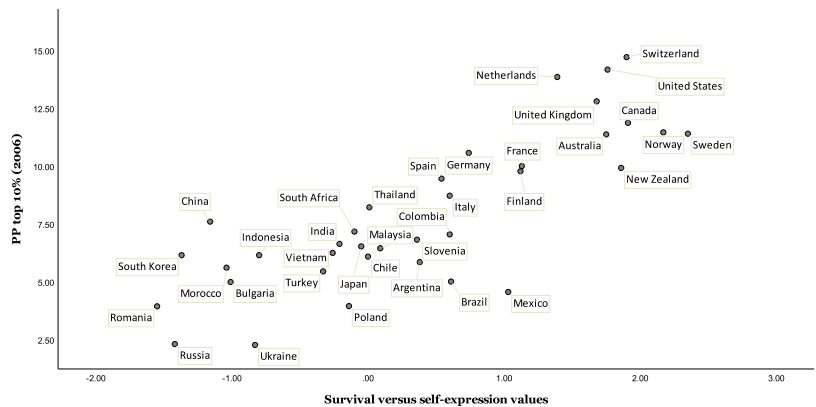

**Fig. 1.** The survival versus self-expression values and citation impact (measured by PP$_{top10\%}$).

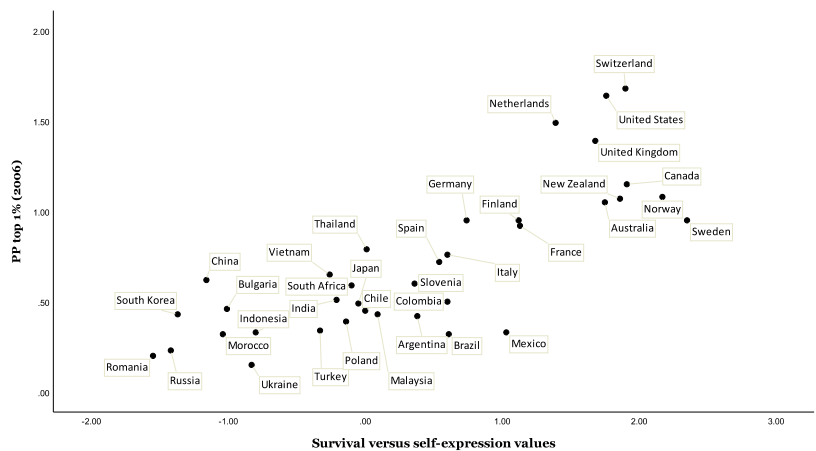

**Fig. 2.** The survival versus self-expression values and citation impact (measured by PP$_{top\ 1\%}$).

### 4.3. Summary of the results

The summary of the statistical analyses is depicted in Table 4. This table also includes results of the linear regression analyses and Pearson's correlation analyses from the supporting information document. As the results in the table reveal, the associations of the cultural dimensions/values with citation impact vary depending on the variables included and methods used.

The fractional logistic regression analyses show that, after controlling for confounding variables, none of HNCD dimensions are statistically significant predictors of citation impact of nations. The Pearson's correlation analyses show a stable correlation of four dimensions of HNCD (power distance, individualism, uncertainty avoidance, and indulgence) and of one dimension of IWCV (survival versus self-expression values) with citation impact before and after controlling for confounding variables. The linear regression analyses reveal that the four noted dimensions of HNCD are not statistically significant predictors of citation impact of nations, and only uncertainty avoidance has a marginal effect on one citation impact indicator (PP$_{top\ 1\%}$).

Thus, the only inclusive result of these statistical analyses is that just one cultural dimension of survival versus self-expression values seems to be a predictor of national citation impact. The finding was stable (robust) before and after controlling for several confounding variables (Table 4).

To better understand the characteristics of this robust relationship, we drew the scatterplots of survival versus self-expression values and PP$_{top10\%}$ and PP$_{top\ 1\%}$ (see Figs. 1 and 2).[1] As it is shown in Fig. 1, most of the Asian, African, and South American nations are scattered in the down-left corner. Those nations have mostly survival values and low citation impact values. European countries with higher scores on self-expression values have higher citation impact values. This pattern is

---
[1] The scatterplots for all other cultural dimensions and citation impact indicators are available in the supporting information (see Sections 1–3, scatterplots).

**Table 4**
Summary of the statistical analyses for the associations of culture and citation impact.

| | | Fractional logistic regression | | | | Pearson's correlation analysis | | | | Linear regression analysis | | | |
|---|---|---|---|---|---|---|---|---|---|---|---|---|---|
| | | Before | | After | | Before | | After | | Before | | After | |
| | | $PP_{top\ 10\%}$ | $PP_{top\ 1\%}$ | $PP_{top\ 10\%}$ | $PP_{top\ 1\%}$ | $PP_{top\ 10\%}$ | $PP_{top\ 1\%}$ | $PP_{top\ 10\%}$ | $PP_{top\ 1\%}$ | $PP_{top\ 10\%}$ | $PP_{top\ 1\%}$ | $PP_{top\ 10\%}$ | $PP_{top\ 1\%}$ |
| **Hofstede's national culture dimensions** | Power distance | −2.9[i] | | | | −.676[ii] | −.615 | −.469 | −.420 | −.301[iii] | | | |
| | Individualism | | .006 | | | .625 | .598 | .363 | .326 | .274 | .299 | | |
| | Masculinity | | | | | | | | | | | | |
| | Uncertainty avoidance | −3.2 | −.008 | | | −.504 | −.548 | −.421 | −.511 | −.324 | −.384 | | −.211 |
| | Long-term orientation | | | | | | | | | | | | |
| | Indulgence | | | | | .363 | .381 | | | | .232 | | |
| **Inglehart-Wetzel's cultural values** | Traditional versus secular-rational values | | | | | | | | | | | | |
| | Survival versus self-expression values | 2.3 | .28 | 1.4 | .18 | .814 | .792 | .625 | .586 | .780 | .759 | .593 | .503 |

Notes for Table 4:
*Before*: the direct link of cultural dimensions/values and citation impact is considered.
*After*: the associations of cultural dimensions/values after controlling for confounding factors.
[I] marginal effects in percentage. [ii] Pearson's correlation coefficients. [iii] $\beta$.
*Confounding variables* include the nation-level data on national self-citations, international co-authorships, investments in research and development, international migrant stock, number of researchers of each nation, language, and productivity.

similar for North American countries (i.e. the United States and Canada). Fig. 2 shows similar patterns for the association of survival versus self-expression values and the citation impact of nations, measured by PP$_{top\ 1\%}$.

Figs. 1 and 2 also show that nations which have common cultural and historical ties have similar patterns of citation impact. This pattern is especially noticeable for European North/Northwest nations (such as Norway, Sweden, and Finland), Anglosphere nations (English-speaking nations such as United States, Canada, Australia, and New Zealand), and the European South/Southeast nations (France, Italy, and Spain) in which self-expression values are predominant and citation impact is high. Additionally, the association of cultural values with science quality is observable for European central East/East nations and Ex-Soviet nations (Bulgaria, Poland, Romania, Russia, and Ukraine), and Asian East/Southeast nations (e.g. China, India, South Korea, and Japan) in which (mostly) survival values are prevalent and citation impact is relatively low.

## 5. Discussion

The purpose of this study was twofold: 1) to investigate the associations of national culture values/dimensions with citation impact, and 2) to see if the links of cultural values/dimensions with citation impact is stable after controlling for confounding variables of national self-citations, international co-authorships, investments in research and development, international migrant stock, number of researchers of each nation, language, and productivity. The results showed that after controlling for the effects of the confounding factors, the associations of HNCD with citation impact disappeared. Only one national culture dimension/value was a stable predictor of citation impact: survival versus self-expression values. The self-expressed nations were more likely to have higher citation impact than the survival nations (and vice versa). This finding was stable before and after controlling for confounding variables based on various statistical analyses.

The stable associations of survival versus self-expression values with citation impact are in accordance with previous research. Inglehart and Welzel (2005) revealed that the "sense of existential security" (i.e. the religion becomes less central when the people reach economic security) and "individual agency" (i.e. people acts on his/her own behalf) are mostly prevalent in nations with secular-rational values and self-expression values. Welzel and Inglehart (2010) further showed that emancipative values – a building block of self-expression values (defined as liberty and equal opportunities for all citizens) – contribute to the general human empowerment process. The possibility to exercise freedom (i.e. democracy) is a strong component of human empowerment (Welzel et al., 2003). In democratic societies, there is higher trust among citizens (Inglehart & Oyserman, 2004), and this trust could serve as a positive accelerator of academic productivity (Chalker & Loosemore, 2016; Dwivedi, 1985), innovativeness (Bavec, 2007), international research collaborations (Bagshaw et al., 2007), and the quality of relationships (Wong & Sohal, 2002). One could expect, therefore, that citizens have more trust (in general) in self-expressed nations which could result in high quality relationships, eagerness to participate in group projects and diverse team groups, and higher quality publications (measured in terms of citation impact).

Although Wagner et al. (2018) showed that academic openness has a strong relationship with national citation impact, it is possible that academic openness is not the main predictor of citation impact. It might act as a mediator in this process, and other national/societal dimensions such as national cultural values affect the degree of openness of nations, and accordingly, citation impact. Allik et al. (2020) showed that "good governance" plays an important role in scientific impact of nations. Good governance (i.e. responsible government, low violence in society, governance of laws, transparent systems, and low corruption) is the main characteristic of self-expressed nations, too. For example, Inglehart and Oyserman (2004) showed that the low level of trust and tolerance was more common in nations with survival values in which the economic and physical securities are on central focus, while societies with self-express values were characterized by the freedom of expression and the personal choice.

Trust is a stimuli for collaborations and innovations of individuals and nations (Doh & Acs, 2010). A low level of trust is associated with a higher level of uncertainty (Möllering, 2005) or uncertainty intolerance (Frederiksen, 2014) of individuals, and a high level of uncertainty is correlated with low productivity (Lazear & Spletzer, 2012). This uncertainty might negatively affect the motives for scholarly communications, too. Furthermore, other research has shown that when the uncertainty level in a country is high, the people are more likely to "develop low cost strategies to manage their worries and concerns" (Alaszewski & Coxon, 2009). Thus, one could speculate that researchers in countries with a low level of trust (among community members) might have pragmatic approaches focusing on *success on the first try* rather than *targeting top journals* in relevant fields to publish their papers. Focusing on *targeting top journals* comes with higher degree of uncertainty for researchers regarding peer review results.

Our findings may point out the embeddedness of scientific performance (measured in terms of citation impact) in national cultures. It adds the often-neglected cultural insight to the science communication system. The findings reveal a stable positive association of self-expression values with strong science. Although cultural dimensions do not normally change in short-term, and they are resultant to many factors, the research system of nations could benefit from academic freedom, diversity, and autonomy of researchers. Our data highlights the importance of European approaches and initiatives to science communication such as the European Research Area (ERA, aimed at free interactions of research, researchers, and technology).

The empirical research described in this paper has several limitations that should be considered in the interpretation of the results:

(1) In this study, we used nations as 'units of analysis' and explored the links of two cultural dimensions/values with citation impact. We controlled for the effects of seven most relevant confounding factors of national self-citations, international co-authorships, investments in research and development, international migrant stock, number of researchers of each nation, language, and productivity. However, the citation impact of nations might be resultant to (many) other possible factors that are not considered here.

(2) Because of their wide disciplinary applicability, we used two nation-level cultural frameworks and related datasets (HNCD and IWCV) in this study. However, there are general limitations of such dimensionalized cultural values. As Taras et al. (2009, p. 362) emphasized, culture is a "complex" construct. While the number of quantitative cultural dimensions have been increasing from four dimensions in early studies of Hofstede to eighteen dimensions in later (GLOBE) studies, "it is still too early to claim that every aspect of culture is captured by any single model or even by all existing models taken together". In order to replicate the findings of this study, or to find the links of other dimensions of culture with research impact, we suggest that future research uses other nation-level cultural datasets such as the seven Schwartz cultural value orientation scores (Schwartz, 2008). These scores are freely available for 80 countries. Another alternative is GLOBE's cultural dataset (House et al., 2004) which includes data for 66 nations.

(3) The citation impact indicators of this study are based on datasets extracted from the Web of Science database, which mostly focuses on English publications. Furthermore, only articles and reviews have been considered (but not books and other document types). The results of this study on the national level might change without these limitations concerning language and document type.

(4) We investigated the associations of two different cultural theories (HNCD and IWCV) with two advanced citation impact indicators ($PP_{top10\%}$ and $PP_{top\ 1\%}$). The study revealed just one cultural bipolar dimension (survival versus self-expression values) as a stable predictor of citation impact of nations (before and after controlling for confounding variables). Since the associations of survival versus self-expression values with citation impact indicators were stable in all parallel statistical analyses (i.e. correlation, linear regression, and fractional logistics regression analyses), it could be regarded as a robust finding. However, the assumptions of HNCD and IWCV (and their included dimensions/values) are not the same and the proposed robustness might not necessarily follow from the results.

(5) In this study, we used field-normalized indicators to control for field effects in citations. Our results are based on the total output of nations. In future studies, our general results might be completed by results focusing on various broad fields. It would be interesting to know whether the links of cultural values and citation impact differ by broad fields or not.

**Author contributions**

**Mahmood Khosrowjerdi:** Conceived and designed the analysis, Collected the data, Contributed data or analysis tools, Performed the analysis, Wrote the paper.
**Lutz Bornmann:** Conceived and designed the analysis, Collected the data, Contributed data or analysis tools, Performed the analysis, Wrote the paper.

**Supporting information**

The supporting information for this study is freely available from the *Inland Norway University (INN) Open Research Data*, see https://dataverse.no/dataverse/inn.

**Acknowledgements**

The bibliometric data used in this paper are from an in-house database developed and maintained by the Max Planck Digital Library (MPDL, Munich) and derived from the Science Citation Index Expanded (SCI-E), Social Sciences Citation Index (SSCI), Arts and Humanities Citation Index (AHCI) prepared by Clarivate Analytics (Philadelphia, Pennsylvania, USA). We would like to thank Jonathan Adams for providing feedback on an earlier version of the paper, the anonymous reviewers for the constructive comments on a previous version of this manuscript, and Alexander Tekles for providing the national self-citation data used in this study.

**Mahmood Khosrowjerdi** is a senior academic librarian (førstebibliotekar) in the Inland Norway University of Applied Sciences, Elverum, Norway. He has PhD in Library and Information Science from the Oslo Metropolitan University. He has published in the Journal of the Association for Information Science and Technology (JASIST), Library and Information Science Research, and the Journal of Documentation, among others. His research interests are socio-cultural studies of information behavior, scientific communications, and research evaluation.

**Lutz Bornmann** is habilitated sociologist of science and works at the Science Policy and Strategy Department in the Administrative Headquarters of the Max Planck Society in Munich (Germany). His research interests include research evaluation, peer review, bibliometrics, and altmetrics. He is member of the editorial board of Quantitative Science Studies (MIT Press), PLOS ONE (Public Library of Science), and Scientometrics (Springer) as well as advisory editorial board member of EMBO Reports (Nature Publishing group). Clarivate Analytics (http://highlycited.com) lists him among the most-highly cited researchers worldwide between 2014 and 2019 . He is recipient of the Derek de Solla Price Memorial Medal in 2019.